\documentclass[10pt,  twoside, onecolumn, notitlepage]{article}
\usepackage[vmarginratio=1:1,a4paper,body={6.5in, 9.5in}]{geometry}

\usepackage{amsmath,amssymb}
\usepackage{graphicx}

\title{Inappropriate use of L-BFGS\\ \large{Illustrated on frame field design}}
\author{Nicolas Ray\\
	INRIA \\
	\and
	Dmitry Sokolov \\
	Universitée de Lorraine \\
	}

\date{\today}
\begin{document}
\maketitle

\begin{abstract}

L-BFGS is a hill climbing method that is guarantied to converge only for convex problems. In computer graphics, it is often used as a black box solver for a more general class of non linear problems, including problems having many local minima.

Some works obtain very nice results by solving such difficult problems with L-BFGS. Surprisingly, the method is able to escape local minima: our interpretation is that the approximation of the Hessian is smoother than the real Hessian, making it possible to evade the local minima.

We analyse the behavior of L-BFGS on the design of $2D$ frame fields. It involves an energy function that is $C^\infty$, strongly non linear and having many local minima. Moreover, the local minima have a clear visual interpretation: they corresponds to differents frame field topologies.

We observe that the performances of LBFGS are almost unpredictables: they are very competitive when the field is sampled on the primal graph, but really poor when they are sampled on the dual graph.

\end{abstract}

\section*{Introduction}

Producing low curvature frame fields with a simple topology is interesting for computer graphics applications like global parameterization, quad remeshing, surface segmentation etc.
The problem is formulated as a minimization problem with an estimation of a smoothed curvature to ease the optimization process.

The resulting non linear optimization problem is usually solved by a dedicated algorithm inspired by the geometric interpretation of a dual problem (producing a smooth unit vector field) \cite{PGP,DFD,NROSY,kowalski2012}.
However, it is also possible \cite{Hertzmann00illustratingsmooth} to directly use an existing non linear solver like L-BFGS.
This approach is also used in \cite{Huang:2011,Li:2012} to extend the frame fields to volumes.

This work aims to illustrate the ability of each approach to find a good quality local minimum of the objective function.
As each minimum corresponds to a frame field topology, we observe minima via their induced frame field topology.

\section{Problem settings}

The optimization problem is completely formalized in \cite[\S 2]{FrameFields}, but can be summarized by the following elements:

\begin{itemize}

\item {\bf Domain representation:} The domain is a $2D$ triangle mesh.
The method optimizes the frame field defined on a graph extracted from the mesh. It can be either the primal graph (vertices / edges) or the dual graph (triangles / dual edges).

\item {\bf Field representation:} A frame is defined on each node $i$ of the graph.
A frame is represented by an angle $\theta_i$, the set of unit vectors is generated as $(\cos(\theta_i+k\pi/2),\sin(\theta_i+k\pi/2))$ with $k\in \{0,1,2,3\}$.

\item {\bf Energy to minimize:} The energy to minimize is the sum of squared distances between adjacent samples.
The distance between two samples $i,j$ is defined as $\sqrt{2(1-\cos(4\theta_i-4\theta_j))}$.
Initially this formulation comes from the representation vector \cite{PGP}, it smoothes the true $C^0$ field curvature.

\item {\bf Boundary condition:} For each node $i$ located on the boundary of the mesh, we estimate the normal of the boundary $\vec{n_i}$ and
force one member vector of the frame to match the normal: $\theta_i = \measuredangle\{\vec{n_i},(1,0)\}$.

\end{itemize}

To summarize, we need to find $\theta$ that minimizes 2$\sum_{ij} \left(1-\cos(4\theta_i-4\theta_j)\right)$ under the constraint $\theta_i = \measuredangle\{\vec{n_i},(1,0)\}$ for boundary samples.

\section{Solvers}

The problem is a highly non linear, but the energy is smooth ($C^{\infty}$) everywhere. We can use existing solvers as a black box (L-BFGS), or define a dedicated solver inspired from the specific geometry of the problem.

\subsection{L-BFGS}

Using L-BFGS requires to find an initial value for $\theta$, and to be able to evaluate the energy and its derivatives. The initial value is directly given by the constraints on the boundary. For other nodes, we compare a random initialization (in the range $[0,\pi/2[$) and values propagated from the constraints. The derivative of the energy with respect to $\theta_i$ are easy to evaluate: $\sum_{ij} 8\sin(4\theta_i-4\theta_j)$ for inner vertices and $0$ for boundary vertices.

\subsection{Dedicated solver}

The problem have an alternative geometric interpretation: it is equivalent to find the smoothest unit vector field defined by $\vec{v_i} =(\cos(4\theta_i), \sin(4\theta_i))$, subject to the same boundary constraints.

Indeed, the squared distance between adjacent vectors is exactly our energy:
\begin{align}
\|\vec{v_i}-\vec{v_j}\| &= (\cos(4\theta_i)-\cos(4\theta_j))^2 +(\sin(4\theta_i)-\sin(4\theta_j))^2\\
  &=(2-2\cos(4\theta_i-4\theta_j))^2
\end{align}

We replace the variables $\theta_i$ by the unit $2D$ vectors $\vec{v_i} = (\cos(4\theta_i),\sin(4\theta_i))$. The optimization problem becomes to minimize $\sum_{ij} (\vec{v_i}-\vec{v_j})^2$ under the boundary constraints, plus the new unit norm constraint $\|\vec{v_i}\|=1$ that is required to invert the change of variable.

All methods using this formulation find an initial solution by relaxing the unit norm constraint and solving the resulting linear problem. Some further optimize the energy, but we only consider the initialization step in this work.

\section{Comparison}

\begin{figure}[h]
\centerline{ \includegraphics[width = 14cm]{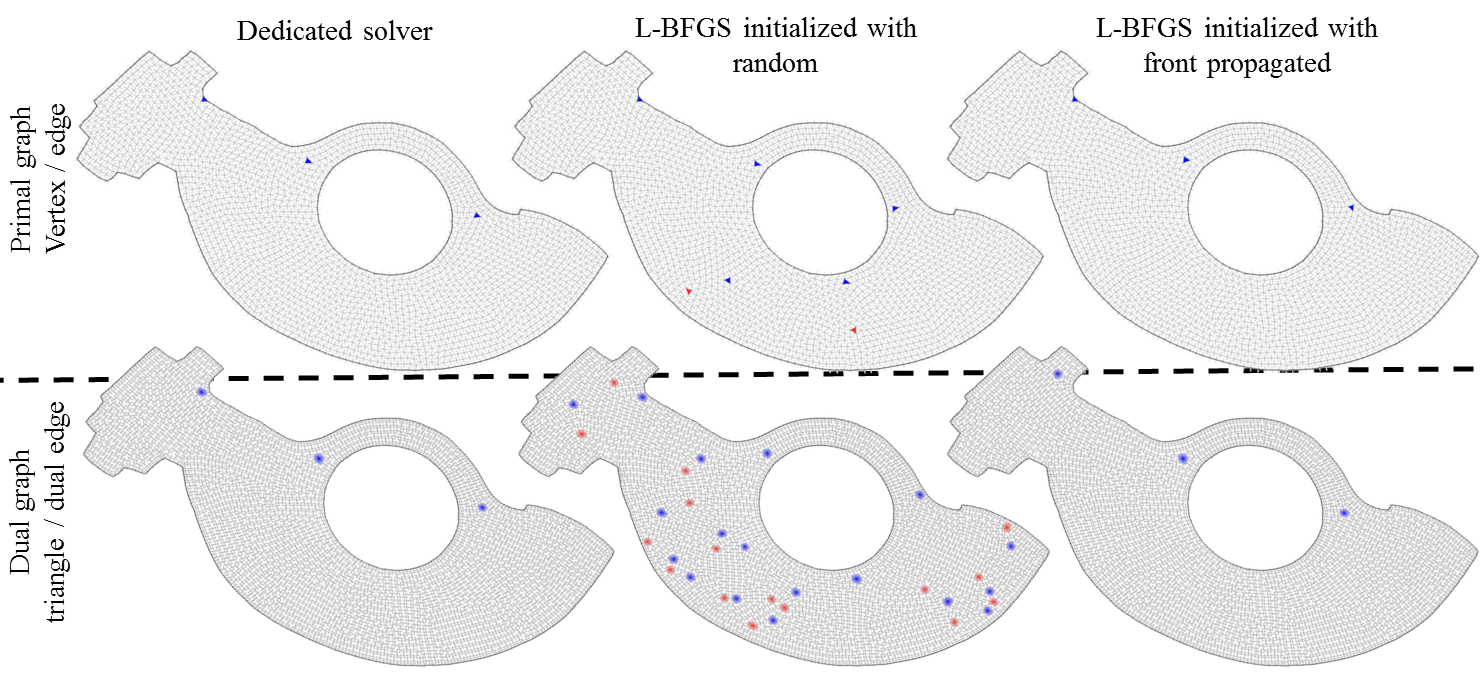}}
\centerline{ \includegraphics[width = 14cm]{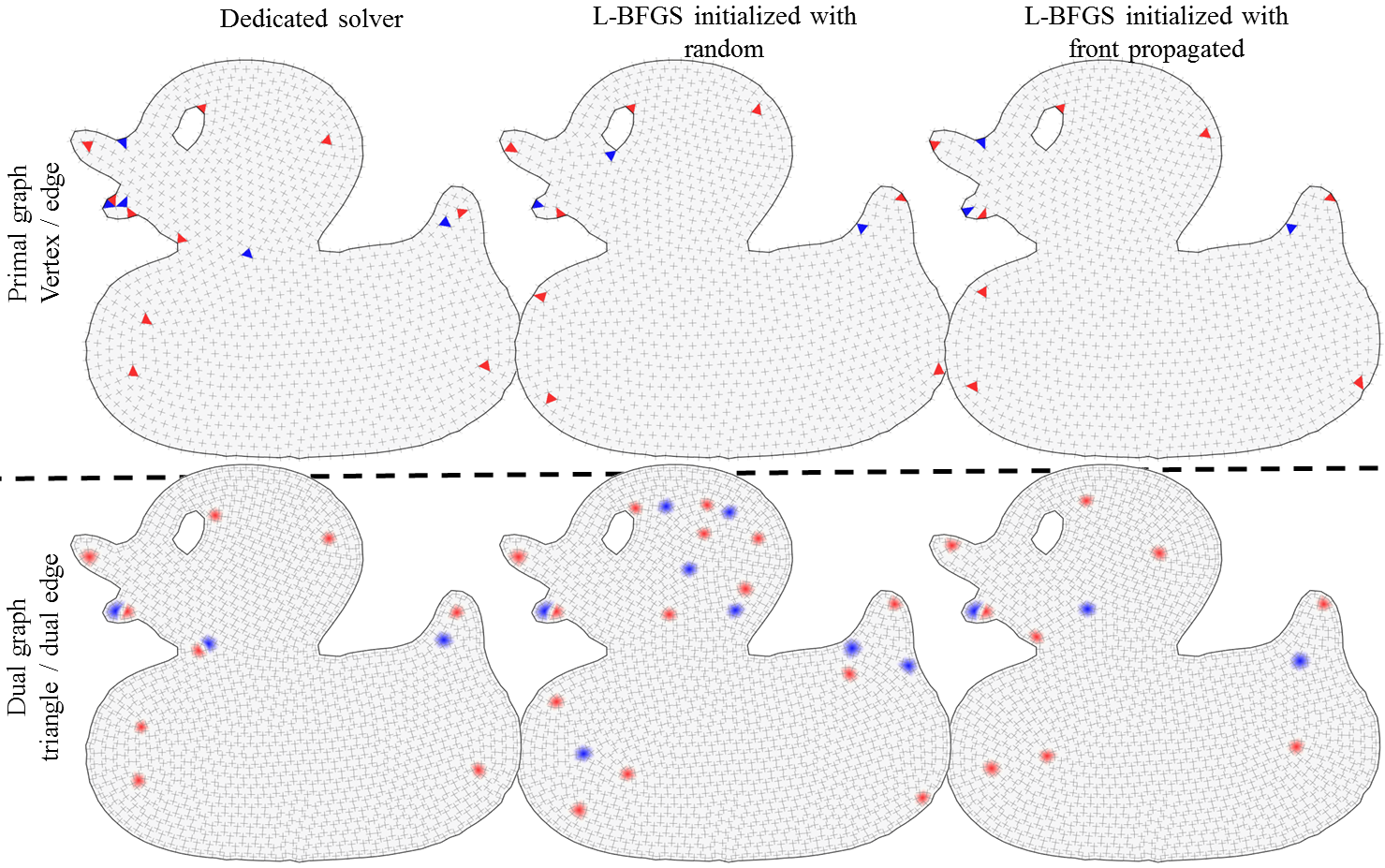}}
\caption{We compare different settings on two models. For each model, the frame fields of the first row are sampled on the (vertex / edge) graph and on the (triangle / dual edge) graph for the second row. The first column is the dedicated algorithm, the second column is L-BFGS with random initialization and the last column is L-BFGS initialized by an advancing front method started from the boundary.}
   \label{rocker}
\end{figure}

\begin{figure}[h]
\centerline{ \includegraphics[width = 14cm]{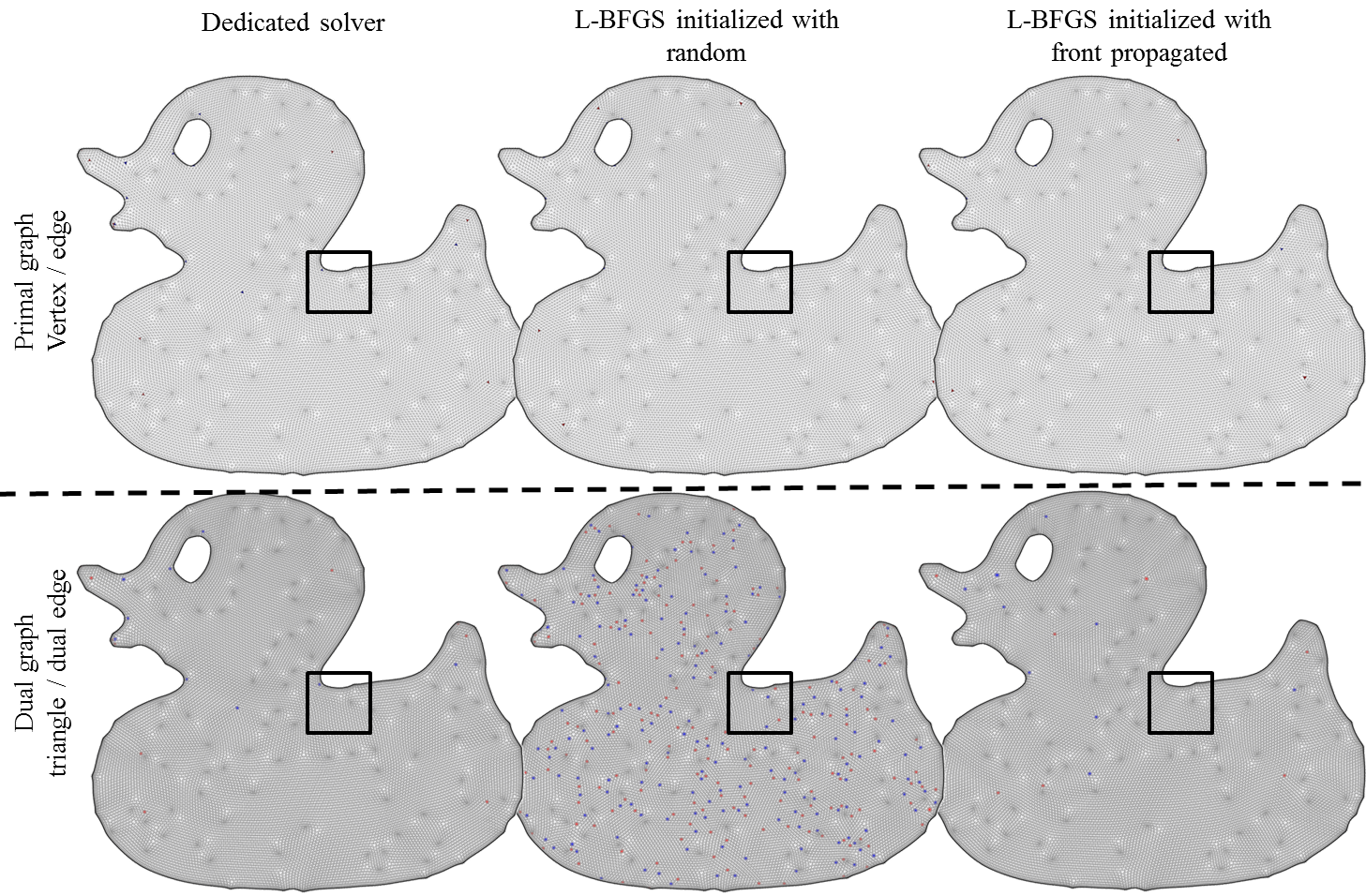}}
\centerline{ \includegraphics[width = 14cm]{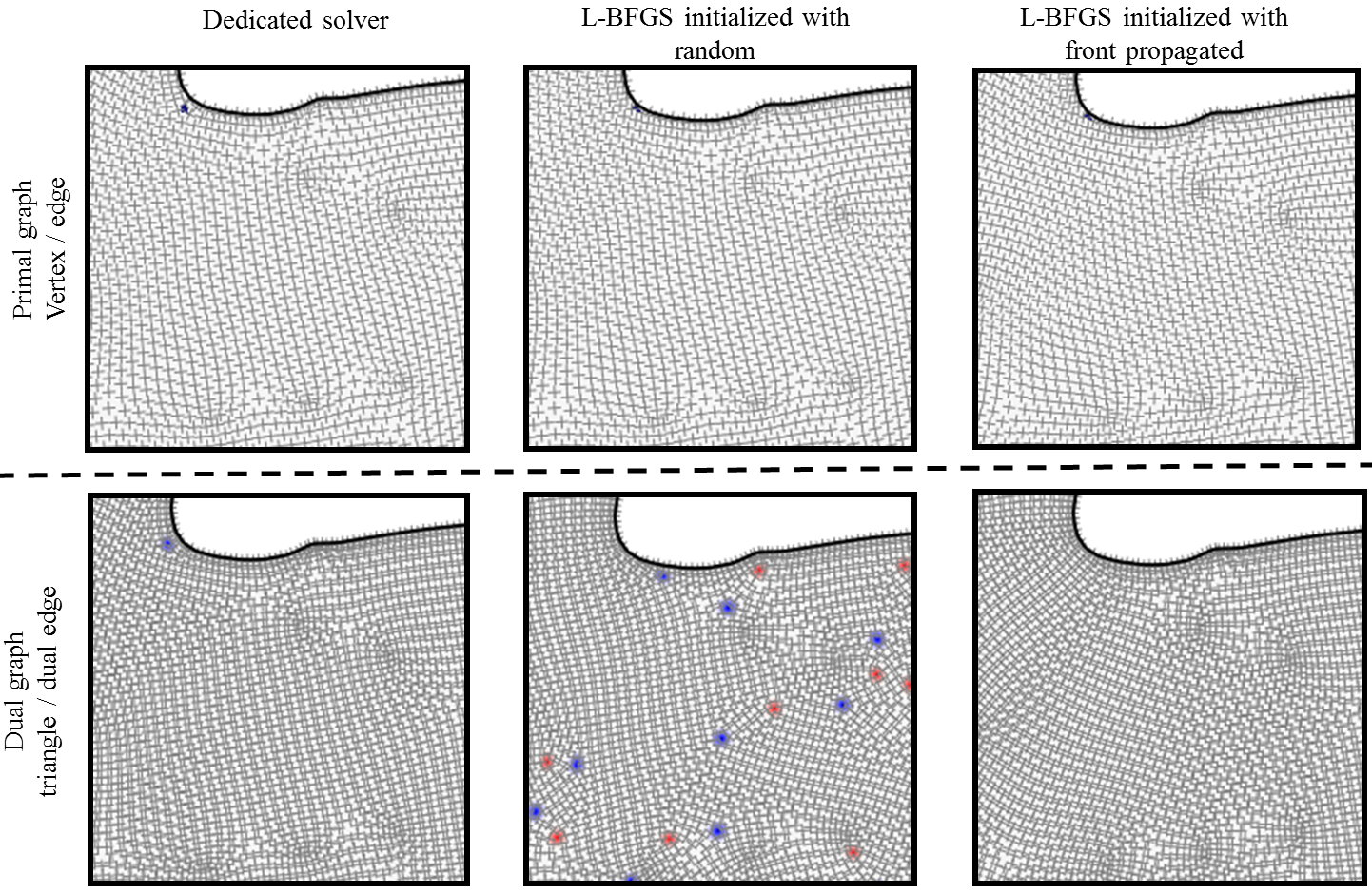}}
\caption{We compare different settings on higher resolution mesh. Images are organized as in Fig~\ref{rocker}}
   \label{duck}
\end{figure}

\begin{figure}[h]
\centerline{ \includegraphics[width = 14cm]{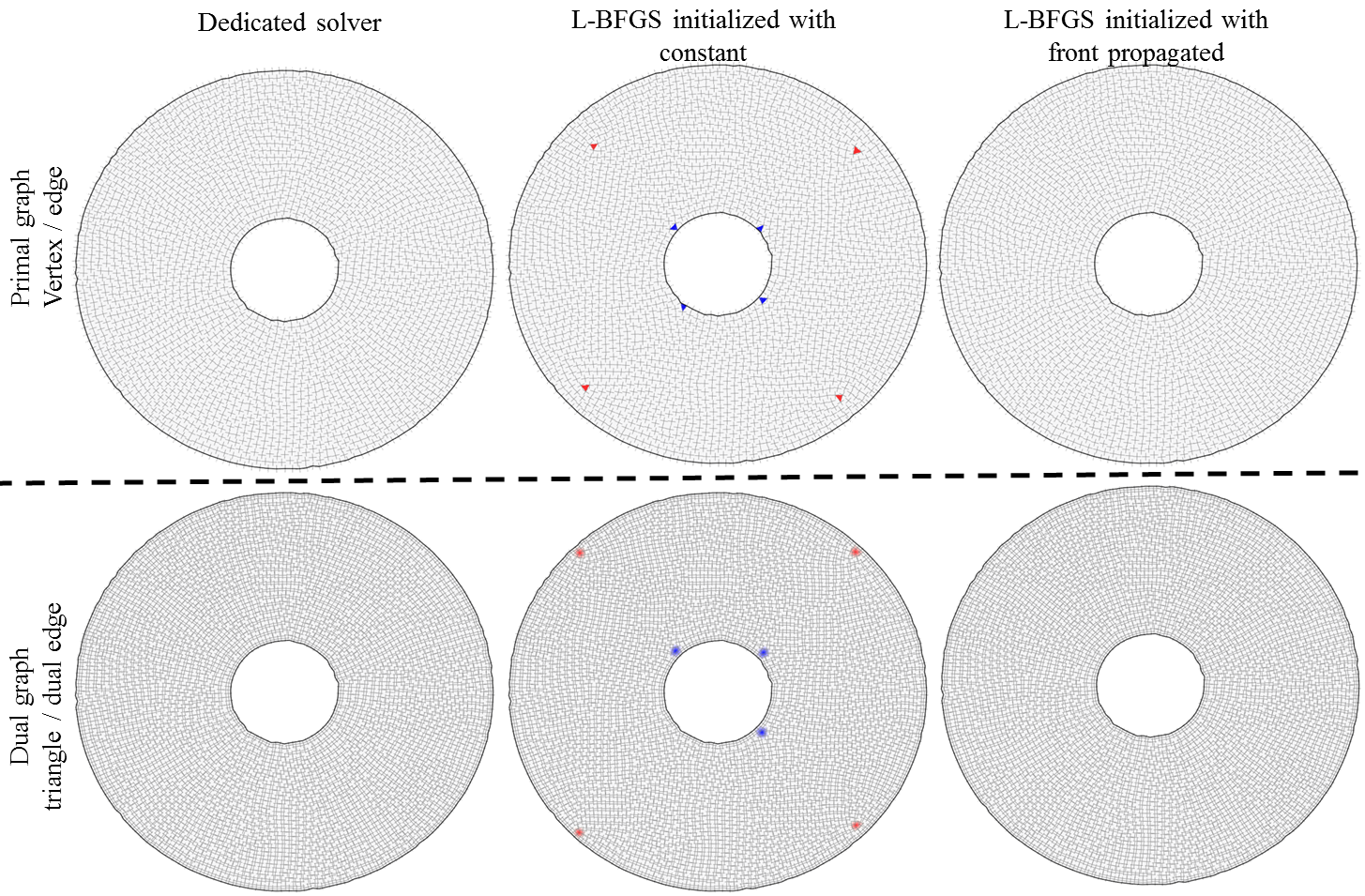}}
\caption{The frame fields of the first row are sampled on the (vertex / edge) graph and on the (triangle / dual edge) graph for the second row. The first column is the dedicated algorithm, the second column is L-BFGS with initialization to $\vec{0}$ and the last column is L-BFGS initialized by an advancing front method started from the boundary.}
   \label{ring}
\end{figure}

Our objective is to compare the quality of the local minima that are found by the different algorithms. A fair geometric interpretation of these local minima is given by the singularities of the frame field. Therefore, we do not consider speed or convergence, but focus only on the observation of the field topology (Fig.~\ref{rocker}, \ref{duck}, and \ref{ring}).

We compare L-BFGS with random initialization, L-BFGS with field initialized by front propagation from boundary, and the dedicated solver. We also test on the primal graph (vertex/edge) and the dual graph (triangle/dual edge).

We observed that :
\begin{itemize}
\item L-BFGS with a good initialization always give results that are similar to the dedicated algorithm. It is not surprising because the initialization have a nice topology.
\item L-BFGS with a field sampled on vertices is able to find an acceptable topology, even when randomly initialized.
\item L-BFGS with a field sampled on triangle is not able to evade the initial local minima. Leading to a complex frame field topology. This difficulty seems directly related to the distance (in number of edges) between singularities that could be simplified, as we can observe on higher resolution mesh  Fig~\ref{duck}.
\item L-BFGS is never able to change the topology that is not due to singularities: as illustrated on Fig~\ref{ring}, the rotation of the field around the hole (that is part of the topology \cite{NSDF}) is too global to be changed by L-BFGS.
\end{itemize}

Our intuitive explanation for these different behaviors with respect to the sampling of the field (graph {\it versus} dual graph) comes from the number of edges of loops in the graph. In vertex / edge graph,  singularities are defined by the sum of curvature along edges around triangles. Therefore, around a singularity, there is only $3$ edges to share the amount of curvature of the singularity ($+/-\pi/2$), so it is quite easy for them to increase/decrease by $\pi/2$ to move the singularity across the edge. In the triangle / dual-edge graph, there is an average of $6$ dual edges sharing the curvature of the singularity, making it more difficult for one of them to change its value.

If such explanations are possible to construct a posteriori, it would be very difficult to predict the behavior of the solver before testing.

\section{Conclusion}
\label{conclusions}

Given the highly non linear and non convex problem of frame field smoothing, L-BFGS works surprisingly well in some cases (when sampled on vertices). However, it fails to find an acceptable minima when sampled on triangles. We discovered that L-BFGS is like magic when used for non convex problems: it makes incredible things, but we can hardly predict how it will behave on a (apparently) similar scenario.

\bibliographystyle{alpha}
\bibliography{periodic_solver}
\end{document}